\documentclass{ws-procs975x65}
\usepackage{latexsym}
\usepackage{amssymb}

\begin{document}
\title{From Black Hole quantization to universal scaling laws}

\author{Salvatore Capozziello${}^\ddag$, Gerardo Cristofano${}^\ddag$, Mariafelicia De Laurentis${}^\ddag$  and Orlando Luongo${}^\ddag{}^\S$}
\address{$^\ddag$Dipartimento di Scienze Fisiche and INFN, Universit\`a di Napoli "Federico
II", Via Cinthia, I-80126, Napoli, Italy;\\
$^\S$Instituto de Ciencias Nucleares, Universidad Nacional Aut\'onoma de M\'exico, AP 70543,
M\'exico DF 04510, Mexico.}

\begin{abstract}
Significative developments on the primordial black hole quantization seem to indicate that the structure formation in the universe behaves under a unified scheme. This leads to the existence of scaling relations, whose validity could offer insights on the process of unification between quantum mechanics and gravity. Encouraging results have been obtained in order to recover the observed magnitudes of angular momenta, peculiar radii and virialized times for large and small structures. In the cosmological regime, we show that it seems possible to infer the magnitude of the cosmological constant in terms of the matter density, in agreement with the observed values.
\end{abstract}

\section{Introduction}

Recently, wide interest has been devoted to investigating the universe dynamics, through its fundamental constituents. Novel evidences show that the description of large scale structure can be seen as a consequence of primordial quantum fluctuations. As a result, a unique unification scheme is achieved regarding  quantum mechanics and gravity \cite{capoz1,capoz2}. It follows that a phenomenological scaling law arises and may reproduce the observational structures to different scales. This shows a possible hidden symmetry, as proposed in several works \cite{sak,suk}. By assuming these recipes, a possible black hole (BH) quantization condition is recovered, whose validity spans from the Planck mass up to the whole universe mass. Soon, we show that a more general scaling law is  derived from the BH quantization. By interpreting it as a universal relation, valid in all the universe epoches, it is possible to reconsider the existence of more complex structures at early and late times. In our work, the answer to this issue is shown to be affirmative.

\section{Quantization procedure and the characteristic length}

To fix the ideas, let us start from a Reisser-Nordstrom BH \cite{cristo1,cristo11}, whose mass is given (in natural units) by $M^2=Q_{e}^{2}+Q_{m}^{2}$. Here,  $Q_e$ and $Q_m$ are the electric and magnetic charge respectively. It is a consolidate result that the Reisser-Nordstrom BHs behave like dyons, leading to the possibility of attractive gravitational forces which compensate the repulsive electric and magnetic interactions. By considering $\phi$ as the dilaton field, and $a$  the axion, equating the BH scalar potential to the one obtained in the conformal field theory, we get $M^2=2Q_eQ_m$. Through the use of the so called Dirac quantization, we obtain that $M_{BH}=\sqrt{n}M_{Pl}$, where $n$ is an integer number, $M_{pl}\equiv\sqrt{\frac{\hbar}{G}}$ the Planck mass and $M_{BH}$ the BH mass. Assuming that at the Planck time charged BHs dominated over the whole universe dynamics, the quantization parameter $n$ is proportional to the BH action. This shows that $M$ depends \emph{linearly} on $M_{Pl}$, i.e., more generally, the quantization scheme is linear in terms of the angular momentum, for rotating BHs. To extend this concept to more general scales, it is possible to infer the validity of the ratio $n_{astro}=\frac{r_{S-astro}}{2\lambda_{Compton-astro}}$, where we assumed the Schwarzschild radius ($r_S$) and the  Compton length ($\lambda_{Compton-astro}$) for any astrophysical structures. It has been shown that such a behavior is compatible with the whole universe evolution, i.e. from small scale structures up to the whole universe. Under these hypotheses, it naturally follows that $J=n\hbar$, or alternatively that the angular momentum of the mass $M$ is given by $J=\frac{G}{c}M^2$, which represents a well known relation of particle physics, extended here in the gravitational regime. In order to show how such a scaling relation for the angular momentum applies to any self-gravitating system, let us write $J=\left(\frac{G M^{2}}{R}\right) \left(\frac{R}{c}\right)=E{\cal T}$. Here, we introduced $E$ as the characteristic gravitational energy of astrophysical structures of "size" $R$, and $T$ the time for virializing \cite{cristo1}. Thus, we assume that the action is: $A \cong E{\cal T}$. The corresponding time-statistical fluctuations, $\tau$ are thus ${\tau}\cong\frac{{\cal T}}{\sqrt{N}}$. In particular, once the number of constituents, for example protons, $N$ is known, one can reach the corresponding typical radius ($R_g$); for example, in the case of a galaxy, we obtain $R_{g} \cong 10^{21} \div 10^{22} \mbox{cm} \simeq
1 \div 10 \mbox{kpc}$, which provides a correct range for the observed magnitude; this can be proved also for other structures, spanning from planetary systems to the whole universe.

\section{Recovering the current value of the cosmological constant}

As shown in Sec. II, the typical size can be obtained up to the whole universe mass. To this end, we want to relate the above results in the cosmological regime. Our purpose becomes then to recover the magnitude of the cosmological constant, $\Lambda$; let us therefore assume the Zeldovich relation; it relates the magnitude of the cosmological constant to the proton mass. Such a numerical agreement seems to be not accidental and it is sometimes referred to as coincidence principle. To better explain it, let us notice the interesting formula relating the universe radius $R_u$ to the Compton wavelength of the proton, i.e. $R_u=\left(\frac{\hbar }{M_p c}\right)^3 \frac{c^3}{G\hbar}$.
By replacing the universe radius $R_u$ with the quantity $\Lambda^{-\frac{1}{2}}$, it is straightforward to find  $\Lambda=\left(\frac{\hbar}{Gc^3}\right)^2\left(\frac{M_p c}{\hbar}\right)^6$, which is however six orders of magnitude greater than its current value. To recover the correct numerical value for $\Lambda$, by considering our approach, it is easy to get for $R_u$ the expression: $R_u=10^3\left(\frac{\hbar}{M_p c}\right)^3\frac{1}{l_{Planck}^2}$. With straightforward algebra, it is possible to recast the value of $\Lambda$ according to \cite{cristo2}
\begin{eqnarray}
\Lambda =\frac{1}{10^6}\left(\frac{\hbar G}{c^3}\right)^2 \left(\frac{M_p c}{\hbar}\right)^6\simeq 10^{-60}cm^{-2}\,,
\end{eqnarray}
which appears to be in a good agreement with the order of  magnitude of the observed $\Lambda$. In fact, evaluating the corresponding energy density $\displaystyle{\rho_\Lambda=\frac{c^2\Lambda}{8\pi G}}$, one obtains $
\rho_\Lambda=\frac{1}{10^6}\left(\frac{GM^2_p}{8\pi \hbar c}\right)\left(\frac{M_p c}{\hbar}\right)^3\simeq10^{-29}\frac{g}{cm^3}$. Alternatively, it is noticeable to introduce the number of protons in the universe and to relate such a number to $\Lambda$; the impressive connection between these expressions is found to be $\rho_\Lambda\simeq \frac{4}{\sqrt{n_p^u}}\left(\frac{M_p}{\frac{4}{3}\pi R_p^3}\right)$, where $R_p$ is the radius of the proton. In particular, the fluctuation factor $\left(\sqrt{n_p^u}\right)^{-1}$ plays a fundamental role for determining the magnitude of $\Lambda$ in terms of the mass density \cite{matter}. From its definition, which is however derived from standard statistical hypotheses, one can address the so-called coincidence problem, concerning the unexpected comparable magnitudes between the dark energy and matter densities today.

\section{Final forecasts}

In this work, we proposed the use of a quantization scheme for primordial BH which accounts for the possibility to describe the structures of the universe in a self consistent way. The development of such a quantization procedure is in agreement with the scaling relation of particle physics between angular momentum and mass; in particular,  assuming that the universe is BH dominated at early times, we were able to show that its evolution is scaling invariant. In fact, the main difference with alternative procedures of quantization lies on showing the validity of such a relation at all scales. To this regard, that opens new insights in the cosmological puzzle, in order to reach a possible connection between quantum mechanics and gravity. In conclusion, we simply found that it is possible to relate the astrophysical structures under a unique scaling law relation. From such considerations, by assuming the fluctuation factor and the proton mass, we infer that $\rho_\Lambda\approx4 \rho_m $, in compatible agreement with the observed value of the matter density.

\end{document}